\documentstyle{article}
\def\be{\begin{equation}}
\def\ee{\end{equation}}
\def\bea{\begin{eqnarray}}
\def\eea{\end{eqnarray}}
\def\beaN{\begin{eqnarray*}}
\def\eeaN{\end{eqnarray*}}
\def\ed{\end{document}}
\def\bit{\begin{itemize}}
\def\eit{\end{itemize}}

\def\sig{\sigma}
\def\Sig{\Sigma}
\def\lam{\lambda}

\def\k{\kappa}
\def\alf{\alpha}

\def\di{\partial}

\def\half{{\textstyle{1 \over 2}}}
\def\~{\tilde}
\def\lag{{\cal L}}
\def\m{\label}
\def\l{\left}
\def\r{\right}

\def\goto{\rightarrow}
\def\Bar{\overline}
\def\const{\rm const}

\begin{document}

\centerline{
 \bf A NOTE ON THE DESER-TEKIN CHARGES}
\medskip

\centerline{\it A.~N.~Petrov}
\medskip

\centerline{Relativistic Astrophysics group, Sternberg Astronomical
institute,} \centerline{Universitetskii pr. 13, Moscow 119992,
RUSSIA}

\medskip
\centerline{ e-mail: anpetrov@rol.ru}

\medskip

\begin{abstract}
Perturbed equations for an arbitrary metric theory of gravity in $D$
dimensions are constructed in the vacuum of this theory. The
nonlinear part together with matter fields are a source for the
linear part and are treated as a total energy-momentum tensor. A
generalized family of conserved currents expressed through
divergences of anti-symmetrical tensor densities (superpotentials)
linear in perturbations is constructed.  The new family generalizes
the Deser and Tekin currents and superpotentials in quadratic
curvature gravity theories generating Killing charges in dS and AdS
vacua. As an example, the mass of the $D$-dimensional Schwarzschild
black hole in an effective AdS spacetime (a solution in the
Einstein-Gauss-Bonnet theory) is examined.
\end{abstract}

In the recent series of the current works \cite{DT1}~-~\cite{DT3}
Deser and Tekin develop a construction of conserved charges for
perturbations about vacua in quadratic gravity theories in $D$
dimensions. They apply the Abbott and Deser \cite{AD} procedure for
description of energy of perturbations in de Sitter (dS) and Anti-de
Sitter (AdS) vacua of cosmological Einstein theory. The perturbed
gravitational equations are presented and transformed as follows.
The l.h.s. of the equations is linear in metric perturbations and
differentially conserved with respect to the background vacuum. The
r.h.s. presents a total, for both metric perturbations and matter
sources, symmetrical energy-momentum tensor. Further the perturbed
equations are contracted with a Killing vector of the background and
are transformed into the form where a total conserved current is
expressed through a divergence of an anti-symmetrical tensor density
(superpotential). This allows to describe a Killing charge in a form
of a surface integral. In \cite{DT1}, quadratic models are studied
for $D=4$. In \cite{DT2}, the energy is defined and calculated in
quadratic curvature gravity theories in arbitrary dimensions. Among
these theories Deser and Tekin study the string-inspired
Einstein-Gauss-Bonnet (EGB) model in detail. In \cite{DT3}, for
$D=3$ the conserved charges in the context of topologically massive
gravity are defined and examined.

The aim of the present letter is to suggest an approach, which
generalizes the Deser and Tekin (DT) constructions. Such
generalizations can be useful for the modern intensive development
of models with branes and their impressive variety (see, e.g.,
review \cite{Rubakov}). Besides, our approach makes it possible to
consider all the DT expressions from an unique point of view in a
more simple and clear light. With regard to the brane models, the
EGB gravity in $D>4$ dimensions has a special and significant
interest (see, e.g., reviews \cite{Nojiri,Der}). {\it Namely}, in
the framework of the EGB gravity the DT charges are reformulated in
the Hamiltonian description \cite{Paddila}; conservation laws for
perturbations are constructed using the canonical N{\oe}ther
procedure \cite{branaP1}~-~\cite{DerKatz}. Therefore we construct
conserved charges in arbitrary metric theories of gravity in $D$
dimensions and on arbitrary curved vacuum backgrounds. Then we apply
the general approach to describe conserved charges on AdS
backgrounds in EGB gravity.  As an illustration we present the mass
of the Schwarzschild EBG-AdS black hole calculated using our
approach. We compare also our formulae with the related DT results
\cite{DT2}. Finally, we discuss the differences that can appear due
to different definitions of metrical perturbations.

We shall follow the approach developed in \cite{PP88} for
constructing perturbed theories. Let the system of fields, set of
tensor densities $Q^A$, be described by the Lagrangian $\lag(Q)$
with arbitrary, but finite order of derivatives of $Q^A$. The
corresponding field equations are ${\delta \lag}/{\delta Q^A} = 0$.
Let us decompose $Q^A$ onto the background (fixed) $\Bar Q^A$ part,
denoted by barred symbols, and the dynamic (perturbed) $q^A$ part:
$Q^A = \Bar Q^A + q^A$. The background fields  satisfy the
background equations ${\delta \Bar \lag}/{\delta \Bar Q^A} = 0$ with
$\Bar \lag = \lag (\Bar Q)$. The perturbed system is now described
by the Lagrangian
 \be
\lag^{pert} = \lag (\Bar Q+q) - q^A \frac{\delta \Bar \lag}{\delta
\Bar Q^A} - \Bar \lag
 \m{lag}
 \ee
instead of the original Lagrangian $\lag(Q)$. The background
equations should not  be taken into account before variation of
$\lag^{pert}$ with respect to $\Bar Q^A$. Using the evident property
${\delta \lag}/{\delta \Bar Q^A} = {\delta \lag}/{\delta q^A}$, the
equations of motion related to the Lagrangian (\ref{lag}) are
presented as
 \be
\frac{\delta \lag^{pert}}{\delta q^A} = \frac{\delta }{\delta \Bar
Q^A}\l[\lag(\Bar Q + q) - \Bar \lag\r] = 0\, .
 \m{PERTeqs1}
 \ee
It is clear that they are equivalent to the equations ${\delta
\lag}/{\delta Q^A} = 0$ if the background equations ${\delta \Bar
\lag}/{\delta \Bar Q^A} = 0$ are taken into account. Defining the
``background current''
 \be
\tau^q_A = \frac{\delta \lag^{pert}}{\delta \Bar Q^A} = \frac{\delta
\lag^{pert}}{\delta q^A} -\frac{\delta }{\delta \Bar Q^A}q^B
\frac{\delta \Bar \lag}{\delta \Bar Q^B}\,
 \m{PERTcurrent}
 \ee
and combining this expression with (\ref{PERTeqs1}) one obtains
another form for equation (\ref{PERTeqs1}):
 \be
\frac{\delta }{\delta \Bar Q^A}q^B \frac{\delta \Bar \lag}{\delta
\Bar Q^B} = - \tau^q_A\,.
 \m{PERTeqs2}
 \ee
The l.h.s. in (\ref{PERTeqs2}) is a linear perturbation of the
expression ${\delta \lag}/{\delta Q^A}$.

Now we assume that the theory of the fields $Q^A$ is an {\it
arbitrary metric} theory of gravity with sources $\Phi^B$ in $D$
dimensions, thus $Q^A = \{g_{\mu\nu},\, \Phi^B\}$. Assume also that
the background is presented by a vacuum of this theory, when $\Bar
Q^A = \{\Bar g_{\mu\nu}\}$ with $\Bar \Phi^B = 0$. Thus only the
decomposition $g_{\mu\nu} = \Bar g_{\mu\nu} + h_{\mu\nu}$ has to be
used.  Below we demonstrate that conserved charges of the system and
their properties can be obtained and described analyzing {\it only}
the terms of the type $\lag_1 \equiv h_{\alf\beta} ({\delta \Bar
\lag}/{\delta \Bar g_{\alf\beta}})$, which corresponds to the
subtracted second term in the Lagrangian (\ref{lag}). As an
important case  we consider explicitly only such theories where
$\lag_1$ has derivatives not higher than of second order, like the
EGB gravity. In principle our results can be repeated when $\lag_1$
has derivatives of higher orders.

Thus, the equation (\ref{PERTeqs2}) is rewritten as ${\delta
\lag_1}/{\delta \Bar g_{\mu\nu}} = - \tau_{h,\Phi}^{\mu\nu}$ or in a
more convenient form
 \be
\frac{2}{\sqrt{-\Bar g}}\frac{\delta \lag_1}{\delta \Bar g_{\mu\nu}}
= T_{h,\Phi}^{\mu\nu}\,
 \m{PERTeqs3}
 \ee
with the standard definition:  $ T^{h,\Phi}_{\mu\nu} \equiv
{2}({\sqrt{-\Bar g}})^{-1}{\delta \lag^{pert}}/{\delta \Bar
g^{\mu\nu}}$ for the symmetrical energy-momentum tensor. Keeping in
mind that $\lag_1$ is the scalar density we follow the standard
technique of constructing covariant differential conservation laws
\cite{LL,Mitz}. The identity ${\pounds}_\xi \lag_1 + \di_\alf
(\xi^\alf\lag_1) \equiv 0 $, where the vector fields $\xi^\alf$ are
arbitrary, transforms to the identity:
 \be
 \Bar \nabla_\mu J^\mu \equiv \di_\mu J^\mu
\equiv 0\, ,\m{nabla-m2}
 \ee
   \be
J^\mu \equiv \lag_1\xi^\mu - 2\xi^\nu h_{\rho\nu} \frac{\delta \Bar
\lag}{\delta \Bar g_{\mu\rho}}  - 2\xi_\nu \frac{\delta
\lag_1}{\delta \Bar g_{\mu\nu}} - \frac{\di\lag_1}{\di \Bar
g_{\rho\sig,\mu\nu}}\Bar \nabla_\nu \l(\pounds_\xi \Bar
g_{\rho\sig}\r) + {\pounds}_\xi \Bar g_{\rho\sig}\Bar \nabla_\nu
 \frac{\di\lag_1}{\di \Bar
g_{\rho\sig,\mu\nu}}\, .
 \label{Jmu}
 \ee
Lie derivatives are defined as
 $
\pounds_\xi Q^A \equiv - \xi^\alf \di_{\alf}Q^A + Q^A|^\alf_\beta
\di_\alf\xi^\beta
 $; the background metric $\Bar g_{\mu\nu}$ rises and lowers all the indexes and
defines the covariant derivatives $\Bar \nabla_\mu$. After taking
into account the background equations and using the Killing vectors
$\Bar \xi^\alf$, for which ${\pounds}_{\Bar \xi} \Bar g_{\rho\sig}
\equiv - 2\Bar \nabla_{(\rho} \Bar \xi_{\sig)} =  0$, the identity
(\ref{nabla-m2}) transforms to
 \be \Bar \nabla_\mu \l(\frac{\delta
\lag_1}{\delta \Bar g_{\mu\nu}}\r) \equiv \Bar \nabla_\mu
\l(\frac{\delta }{\delta \Bar g_{\mu\nu}}h_{\alf\beta} \frac{\delta
\Bar \lag}{\delta \Bar g_{\alf\beta}}\r) \equiv 0\, ,
 \m{nabla-m1}
 \ee
and the equations (\ref{PERTeqs3}) give  $\Bar \nabla_\mu
T_{h,\Phi}^{\mu\nu}= 0$ repeating in another form the results from
\cite{DT1, DT2}.

Because equation (\ref{nabla-m2}) is an identity, the expression
(\ref{Jmu}) must be representable as a divergence of an
antisymmetrical tensor density (superpotential) $I^{\mu\nu}$, for
which $\di_{\mu\nu} I^{\mu\nu} \equiv 0$, that is $- J^\mu  \equiv
\Bar \nabla_\nu I^{\mu\nu} \equiv \di_\nu I^{\mu\nu}$ (we use the
sign convention of \cite{Petrov04}). The general expression for
superpotentials initiated by the N{\oe}ther procedure applied to
Lagrangians with derivatives up to a second order and with a
background metric included as an {\it external} metric was obtained
in \cite{Petrov04}. Quadratic gravity is an example of  such a
theory. Thus, for $\lag = \lag(Q;\, \di_\mu Q;\, \di_{\mu\nu}
Q)\equiv \lag_{ext}(Q;\, \Bar \nabla_\mu Q;\, \Bar \nabla_{\mu\nu}
Q)$ the superpotential is
 \be
I_{ext}^{\mu\nu}  = \l(m_\sig{}^{\nu\mu} + \Bar \nabla_\lam
n_\sig^{\lam\nu\mu}\r) \xi^\sig + {\textstyle \frac{2}{3}}\xi^\sig
\Bar \nabla_\lam n_\sig^{[\mu\nu]\lam} - {\textstyle
\frac{4}{3}}n_\sig^{[\mu\nu]\lam}\Bar \nabla_\lam  \xi^\sig\,;
 \m{Jmunu}
 \ee
 \bea
  m^{\alf\beta}_\sig  & \equiv & \l[\frac{\di \lag_{ext}}
{\di  (\Bar \nabla_\alf Q^A)} - \Bar \nabla_\gamma \l(\frac{\di
\lag_{ext}}
  {\di (\Bar \nabla_{\alf\gamma} Q^A)}\r)\r]
 \l. Q^A \r|^\beta_\sig +
 \frac{\di \lag_{ext}}
 {\di (\Bar \nabla_{\alf\gamma} Q^A)}
 \l[ \Bar \nabla_{\gamma}\l(\l.Q^A\r|^\beta_\sig\r) -\delta^\beta_\gamma
\Bar \nabla_{\sig} Q^A \r], \nonumber\\
 n^{\alf\beta\gamma}_\sig & \equiv & \frac{1}{2}
\l( \frac{\di \lag_{ext}}
 {\di (\Bar \nabla_{\alf\gamma} Q^A)}
 \l.Q^A\r|^\beta_\sig + \frac{\di \lag_{ext}}
 {\di (\Bar \nabla_{\alf\beta} Q^A)}
 \l.Q^A\r|^\gamma_\sig \r)\,
 \m{mnL}
 \eea
with the notation $ Q^A |^\beta_\alf$ included in the definition of
$\pounds_\xi Q^A$. The superpotential (\ref{Jmunu}) is
antisymmetrical in $\mu$ and $\nu$, because $m_\sig{}^{(\nu\mu)} +
\Bar \nabla_\lam n_\sig^{\lam(\nu\mu)} \equiv 0$ \cite{Petrov04}.

Now let us adopt the expression (\ref{Jmunu}) to the Lagrangian
$\lag_1$. We set $\Bar g_{\mu\nu} \goto g_{\mu\nu}$ and construct
the coefficients (\ref{mnL}) with $\lag_1 =  \lag_1(Q;\, \di_\mu
Q;\, \di_{\mu\nu} Q)\equiv \lag_1^{ext}(Q;\, \Bar \nabla_\mu Q;\,
\Bar \nabla_{\mu\nu} Q)$, where $Q^A = \{h_{\mu\nu},\, g_{\mu\nu}
\}$. Then we go back to $ g_{\mu\nu} \goto \Bar g_{\mu\nu}$ and
obtain simple expressions
 \be
m_{1\sig}{}^{\mu\nu}  =  2 \Bar \nabla_\lam \l(\frac{\di \lag_1}{\di
\Bar g_{\rho\nu,\mu\lam}}\r)\Bar g_{\rho\sig}\, ,\qquad
n_{1\sig}^{\lam\mu\nu}  =  - 2 \frac{\di \lag_1}{\di \Bar
g_{\rho(\mu,\nu)\lam}}\Bar g_{\rho\sig}\, .
 \m{mnL1}
 \ee
Substituting these into (\ref{Jmunu}) we can define the
superpotential $I^{\mu\nu}$.

Using the background equations and  the field equations
(\ref{PERTeqs3}), the equations (\ref{nabla-m2}) and  (\ref{Jmu})
define the current, $J^\mu \goto I^\mu$, which is conserved
differentially:
 \be
 \Bar \nabla_\mu I^\mu \equiv
\di_\mu I^\mu = 0\, ,
 \m{nablacurrent}
 \ee
 \be
I^\mu  \equiv  -\sqrt{-\Bar g}T_{h,\Phi}^{\mu\nu} \xi_\nu -
\frac{\di \lag_1}{\di \Bar g_{\rho\sig,\mu\nu}}\Bar \nabla_\nu
\l(\pounds_\xi \Bar g_{\rho\sig}\r) + {\pounds}_\xi \Bar
g_{\rho\sig}\Bar \nabla_\nu
 \frac{\di\lag_1}{\di \Bar
g_{\rho\sig,\mu\nu}}\, .
 \m{current}
 \ee
Besides, for this current one can write an equality: $- I^\mu =
\di_\nu I^{\mu\nu}$, from which (\ref{nablacurrent}) follows
directly. Thus, the differential conservation law
(\ref{nablacurrent}) allows us to construct the conserved charges:
 \be
Q(\xi) = \int_\Sigma d^{D-1} x\l[-I^0(\xi)\r] = \oint_{\di\Sigma}
dS_i I^{0i}(\xi)\,
 \m{charges}
 \ee
where $\Sigma$ is a spatial $(D-1)$ hypersurface $x^0 = \const$ and
$\di\Sigma$ is its  $(D-2)$ dimensional boundary.

The conservation law (\ref{nablacurrent}) and the charges
(\ref{charges}) were constructed for arbitrary vectors $\xi^\mu$,
not only for the Killing ones. This can be useful and important.
Thus, in \cite{PK}, the conservation laws of the type  $- I^\mu =
\di_\nu I^{\mu\nu}$ with {\it conformal} Killing vectors (not only
with Killing ones) of the Friedmann-Robertson-Walker spacetime are
used for constructing so-called integral constraints for
cosmological perturbations. Also, in \cite{PK}, the  use of other
non-Killing vectors is discussed, possibilities of interpretation of
corresponding conservation laws and charges are given with related
references. For the Killing vectors $\xi_\nu = \Bar \xi_\nu$ the
quantities (\ref{charges}) are called as Killing charges. For
$\xi_\nu = \Bar \xi_\nu$ the definition (\ref{current}) gives
$~-I^0(\Bar \xi) = \sqrt{-\Bar g} T_{h,\Phi}^{0\nu}\Bar \xi_\nu$
(compare with \cite{DT2}), whereas the form of $I^{\mu\nu}(\Bar
\xi)$ is the same as of $I^{\mu\nu}(\xi)$. Note that charge scales
can always be absorbed by Killing scaling, that is for $Q(\Bar \xi)$
from (\ref{charges}) ${\cal C}Q(\Bar \xi)$ is also a Killing charge
with ${\cal C}$ constant \cite{Desercom}.

Now let us apply the presented approach to the EGB gravity. The
action of the Einstein theory with a bare cosmological term
$\Lambda_0$ corrected by the Gauss-Bonnet term is
 \bea
 S & = & \int d^D x\l\{\lag_{EGB} + {\cal L}_m\r\}\nonumber\\ & =&
 \int d^D x\l\{-\frac{\sqrt{-g}}{2} \l[{\k}^{-1}\l(R - 2\Lambda_0\r) +
 \gamma\l(R^2_{\mu\nu\rho\sig} - 4 R^2_{\mu\nu} + R^2\r)\r] + {\cal L}_m\r\}
 \,
 \m{EGBaction}
 \eea
where $\k = 2\Omega_{D-2}G_D> 0$  and $\gamma >0$; $G_D$ is the
$D$-dimension Newton's constant. Because the DT analysis \cite{DT2}
was restricted to $\Lambda_0= 0$, we note that we consider a more
general situation. The equations of motion that follow from
(\ref{EGBaction}) are
 \bea
 {\cal E}^{\mu\nu} &\equiv &
  \frac{\delta}{\delta
g_{\mu\nu}}\l(\lag_{EGB}
 + \lag^m\r)  \nonumber\\
 &{\equiv}& \frac{\sqrt{-g}}{2}\l\{\frac{1}{\k}\l(R^{\mu\nu} - \half g^{\mu\nu}
 R + g^{\mu\nu}\Lambda_0\r)\r.\nonumber\\ &+& \l.2\gamma\l[RR^{\mu\nu} -
 2 R^{\mu}{}_{\sig}{}^\nu{}_{\rho} R^{\sig\rho} +
 R^{\mu}{}_{\sig\rho\tau}R^{\sig\nu\rho\tau} - 2 R^{\mu}{}_{\sig}
 R^{\sig\nu}  \r. \r.\nonumber\\ &  - & \l.\l.{\textstyle \frac{1}{4}} g^{\mu\nu}
 \l(R^2_{\tau\lam\rho\sig} - 4 R^2_{\rho\sig} + R^2\r)\r] -
 \tau_m^{\mu\nu}\r\} = 0\,
 \m{EGBequations}
 \eea
where $\tau_m^{\mu\nu}$ is the symmetrical matter energy-momentum
tensor. In the vacuum case, AdS  background is a solution of the
equations (\ref{EGBequations}). Then equations (\ref{EGBequations})
are transformed into
 \be
 \Bar R_{\mu\nu} - \half\Bar g_{\mu\nu}\Bar R + \Lambda_{eff} \Bar
 g_{\mu\nu} = 0\,;
 \m{AdSequations}
 \ee
 \be \Bar R_{\mu\alf\nu\beta} = 2\Lambda_{eff}\frac{(\Bar
g_{\mu\nu}\Bar g_{\alf\beta} - \Bar g_{\mu\beta}\Bar
g_{\nu\alf})}{(D-2)(D-1)},\qquad \Bar R_{\mu\nu} =
2\Lambda_{eff}\frac{\Bar g_{\mu\nu}}{D-2},\qquad \Bar R =
2\Lambda_{eff}\frac{D}{D-2}\, . \label{R}
 \ee
The effective cosmological constant (see \cite{DT2}):
 \be
 \Lambda_{eff} = \frac{\Lambda_{EGB}}{2} \l(1\pm \sqrt{1 -
 \frac{4\Lambda_{0}}{\Lambda_{EGB}}}\r)\,
 \label{Leff+}
 \ee
is the solution of the equation $
 \Lambda^2_{eff} - \Lambda_{eff}\Lambda_{EGB} +
 \Lambda_{EGB}\Lambda_{0} = 0$, where

\noindent $ \Lambda_{EGB} = -
{(D-2)(D-1)}/[{{2\k\gamma}(D-4)(D-3)}]\,$ is defined {\em only} by
the Gauss-Bonnet term.

Let us turn to the equations (\ref{PERTeqs3}) in the framework of
the EGB theory. We set  $\Bar \lag = \Bar \lag_{EGB}$  and obtain
 \be
 {\cal G}_{L-EGB}^{\mu\nu} \equiv
\mp \sqrt{1 -
 \frac{4\Lambda_{0}}{\Lambda_{EGB}}}~~ {\cal G}_L^{\mu\nu} = \k
 T^{\mu\nu}\, ,
 \m{GL=T}
 \ee
i.e. the perturbed (with respect to the vacuum AdS solution) the
equations (\ref{EGBequations}). On the l.h.s. of (\ref{GL=T})
 \bea
2{\cal G}_L^{\mu\nu} \equiv & - &  \Box h^{\mu\nu} - \Bar
\nabla^{\mu\nu} h + \Bar \nabla_\sig
 \Bar \nabla^\nu h^{\sig\mu} + \Bar \nabla_\sig
 \Bar \nabla^\mu h^{\sig\nu} - \frac{4\Lambda_{eff}}{D-2}h^{\mu\nu}
 \nonumber\\
  &-&  \Bar g^{\mu\nu}\l(- \Box h + \Bar \nabla_{\sig\rho}
 h^{\sig\rho} - \frac{2\Lambda_{eff}}{D-2}h\r)
 \m{GL}
 \eea
where $h \equiv h_{\mu\nu}\Bar g^{\mu\nu}$, $\Bar \nabla_{\mu\nu}
\equiv \Bar \nabla_\nu\Bar \nabla_\mu$ and $\Box \equiv \Bar
g^{\mu\nu}\Bar \nabla_\mu\Bar \nabla_\nu$. Under variation in
(\ref{PERTeqs3}) we used  $\lag_1 = \lag_1^{EGB} = h_{\alf\beta}
({\delta \Bar \lag_{EGB}}/{\delta \Bar g_{\alf\beta}}) =
h_{\alf\beta}\Bar {\cal E}^{\alf\beta}$, where $\Bar {\cal
E}^{\alf\beta}$ is the barred expression (\ref{EGBequations}). On
the r.h.s. of (\ref{GL=T}) the symmetrical energy-momentum tensor
$T^{\mu\nu}$ consists of all the non-linear in $h_{\mu\nu}$ terms
and components of $\tau_m^{\mu\nu}$. Reformulating the identity
(\ref{nabla-m1}) for the EGB case one obtains an identical
conservation for the l.h.s. of (\ref{GL=T}): $
 \Bar \nabla_\mu{\cal G}_{L-EGB}^{\mu\nu} \equiv 0$, that naturally
 gives for (\ref{GL=T}): $\Bar \nabla_\mu T^{\mu\nu} = 0$
 (compare with \cite{DT2}).

Constructing the charges for the EGB system we set
$\lag_1=\lag^{EGB}_1$ in (\ref{mnL1}) and obtain
 \bea
 m_{1\sig}{}^{\mu\nu}  &= & \mp \frac{\sqrt{-\Bar g}}{\k}
 \sqrt{1 -
 \frac{4\Lambda_{0}}{\Lambda_{EGB}}}
\Bar g_{\sig\rho} \Bar \nabla_\lam H^{\mu(\nu\rho)\lam},\qquad
 n_{1\sig}^{\rho\mu\nu}  =  \mp \frac{\sqrt{-\Bar g}}{\k} \sqrt{1 -
 \frac{4\Lambda_{0}}{\Lambda_{EGB}}}\Bar
g_{\sig(\lam}\delta^{(\mu}_{\pi)}H^{\nu)\pi\rho\lam};\nonumber\\
 H^{\mu\nu\rho\lam} & \equiv &
 h^{\mu\rho} \Bar
g^{\nu\lam} + h^{\lam\nu} \Bar g^{\rho\mu} - h^{\mu\lam} \Bar
g^{\rho\nu} - h^{\rho\nu} \Bar g^{\mu\lam} +h\l( \Bar g^{\mu\lam}
\Bar g^{\rho\nu} - \Bar g^{\mu\rho} \Bar g^{\nu\lam}\r)\, . \m{mnL2}
 \eea
The substitution of the quantities (\ref{mnL2}) into the expression
(\ref{Jmunu})  gives the superpotential  related to the general EGB
case:
 \be
 I_{EGB}^{\mu\rho} \equiv \pm \frac{\sqrt{-\Bar g}}{\k}
\sqrt{1 -
 \frac{4\Lambda_{0}}{\Lambda_{EGB}}}
 \l(\xi^{[\mu} \Bar \nabla_{\nu}h^{\rho]\nu}  - \xi_\nu
\Bar \nabla^{[\mu}h^{\rho]\nu} - \xi^{[\mu} \Bar \nabla^{\rho]}h -
h^{\nu[\mu}\Bar \nabla^{\rho]}\xi_\nu - \half h \Bar \nabla^{[\mu}
\xi^{\rho]}\r)\, .
 \m{DTsuperpotential}
 \ee
It is expressed through the Abbott-Deser superpotential in the
Einstein theory \cite{DT2,AD}, $I_{AD}^{\mu\rho}$, as
$I_{EGB}^{\mu\rho} = \mp \sqrt{1 - {4\Lambda_{0}}/{\Lambda_{EGB}}}~
I_{AD}^{\mu\rho}$. Thus charges (\ref{charges}) corresponding to
$I_{EGB}^{\mu\rho}$ are rewritten as
 \be
 Q(\xi) = \mp
\sqrt{1 - \frac{4\Lambda_{0}}{\Lambda_{EGB}}}\oint_{\di\Sig} dS_i
I_{AD}^{0i}(\xi) \,.
 \label{31bis}
 \ee

To compare the derivation of perturbations on AdS backgrounds in
(\ref{GL=T}) - (\ref{31bis}) with the DT presentation in \cite{DT2}
one has to set $\Lambda_0 = 0$. Then (\ref{GL=T}) and (\ref{31bis})
with the upper sign, ``$-$'', coincide with corresponding
expressions in \cite{DT2}. The lower sign, ``+'', in (\ref{GL=T})
and (\ref{31bis}) corresponds to the case of the Einstein theory. We
consider both branches together, ``+'' and ``$-$'', as an unique
expression in the frame of the EGB theory. Thus, even for $\Lambda_0
= 0$ the picture (\ref{GL=T}) - (\ref{31bis}) generalizes the DT
results \cite{DT2}. Continuing the comparison, we accent also that
the united scheme gives a simple way for constructing
superpotentials without cumbersome calculations in linear
expressions contracted with Killing vectors.

Let us examine the presented results considering the
Schwarzschild-AdS solution \cite{DT2,BD+}:
 \be
 ds^2 = g_{00} d t^2 + g_{rr} d r^2 + r^2 d \Omega_{D-2}\, ,
 \label{S-AdS}
 \ee
 \be
 -g_{00} = 1 + \frac{r^2}{2\k\gamma (D-3)(D-4)}
\l\{1 \pm \sqrt{1 - \frac{4\Lambda_{0}}{\Lambda_{EGB}} + 4\k\gamma
(D-3)(D-4)\frac{r_0^{D-3}}{r^{D-1}}} \r\}\, ,
 \label{39bis}
 \ee
$-g_{00} =  g^{-1}_{rr}$. In linear approximation (\ref{39bis})
gives
 \be
 g_{00} = -\l(1 - r^2\frac{2\Lambda_{eff}}{(D-1)(D-2)}
\r) \mp \l(\sqrt{1 - \frac{4\Lambda_{0}}{\Lambda_{EGB}}}\r)^{-1}
\l(\frac{r_0}{r}\r)^{D-3}\, .
 \label{39bis++}
 \ee
Thus for ``$\pm$'' in (\ref{Leff+}) we have ``$\mp$'' in
perturbations
 \be
h_{00} \approx h_{rr} \approx \mp \l(\sqrt{1 -
\frac{4\Lambda_{0}}{\Lambda_{EGB}}}\r)^{-1}
\l(\frac{r_0}{r}\r)^{D-3}\,
 \label{h}
 \ee
with respect to AdS background with $\Lambda_{eff}$. For all the
cases of perturbations in (\ref{h}), the formula (\ref{31bis}) with
the time-like vector $\Bar \xi^\mu = (-1,\, {\bf 0})$ gives the
total conserved energy of the Schwarzschild-AdS solution:
 \be
 E = \frac{(D-2)r_0^{D-3}}{4G_D}\, ,
 \label{E}
 \ee
which is the standard accepted result (see \cite{DT2,DerKatzOgushi}
and references therein). We note that recently Deser with co-authors
have reached similar conclusions \cite{Desercom}.

Let us discuss the case  $\Lambda_{0}= 0$. The form of perturbations
(\ref{h}) is reduced to $h_{00} \approx h_{rr} \approx \mp
\l({r_0}/{r}\r)^{D-3}$ where ``+'' corresponds to asymptotically
Schwarzschild solution on a flat background, not on AdS one. Then,
at first sight it seems that the formulae of pure Einstein theory,
like Abbott-Deser superpotential or an ADM analog \cite{DT2,AD}, can
be used. However, by including in the action, right from the
beginning, a bare cosmological term $\Lambda_0$, we arrive at an
unique expression for the mass of EGB black hole (\ref{E}), valid
for both branches, ``+'' and ``$-$'' signs in (\ref{39bis}), and for
$\Lambda_0=0$ or $\Lambda_0 \neq 0$. In order to get (\ref{E}) there
is therefore no need to use, as Deser and Tekin in the work
\cite{DT2} do when $\Lambda_0=0$, two different formulae: the
standard ADM formula derived in Einstein's theory for the ``$-$''
sign in (\ref{39bis}) and another one derived in EGB theory for the
``+'' sign.

As a concluding remark it is important to discuss the following
generalization. It is possible to use different definitions for the
metric perturbations, like $h_{\mu\nu} = g_{\mu\nu} - \Bar
g_{\mu\nu}$, $l^{\mu\nu} = \sqrt{- g}g^{\mu\nu} - \sqrt{- \Bar
g}\Bar g^{\mu\nu}$, {\it etc.} After including the notations: $g_a =
\l\{ g_{\mu\nu},\,\sqrt{- g}g^{\mu\nu},\, \ldots \r\}$; $\Bar g_a =
\l\{ \Bar g_{\mu\nu},\,\sqrt{- \Bar g}\Bar g^{\mu\nu},\, \ldots
\r\}$; $h_a = \l\{ h_{\mu\nu},\,l^{\mu\nu},\, \ldots \r\}$ {\em
independent} perturbations (variables): $h^{(a)}_{\alf\beta} \equiv
h_a (\di \Bar g_{\alf\beta}/\di \Bar g_a)$ are defined. With taking
into account the background equations, the forms of expressions in
equations (\ref{PERTeqs3}), (\ref{mnL1}) and, consequently, in the
surface integral (\ref{charges}) are not changed. Only
$h_{\alf\beta}$ is changed by $h^{(a)}_{\alf\beta}$. However, for
different $h_{a}$ the quantities $h^{(a)}_{\alf\beta}$ differ one
from another beginning from the second order. Clearly, these
differences appear on the l.h.s. of (\ref{PERTeqs3}) and,
respectively, in energy-momentum tensors on the r.h.s. of
(\ref{PERTeqs3}), which was noted for the first in \cite{BD} for
perturbations in GR on Minkowski background. Of course, such
differences appear in superpotentials, which are linear in
$h^{(a)}_{\alf\beta}$ (see also a discussion in \cite{Petrov04a}).

In the Einstein gravity with $D=4$, GR, {\em all} the
superpotentials of the above general family on {\em arbitrary}
vacuum backgrounds have the form of the Abbott-Deser superpotential
\cite{AD} by changing only $h_{\alf\beta}$ with
$h^{(a)}_{\alf\beta}$, and  $\Bar \xi^\alf$ with $\xi^\alf$.
Superpotentials of this family also differ from each other starting
from the second order for different $h_a$. This becomes important
for calculations at null infinity as was analyzed by Bondi, van der
Burg and Metzner \cite{BMS}. As a confirmation, it was shown in
\cite{PK} that the Bondi-Sachs momentum calculated with the use of
the Abbott-Deser type superpotential is correct  if the
perturbations $l^{\mu\nu} = \sqrt{- g}g^{\mu\nu} - \sqrt{- \Bar
g}\Bar g^{\mu\nu}$ are used, and not $h_{\mu\nu} = g_{\mu\nu} - \Bar
g_{\mu\nu}$ as in the original Abbott-Deser  superpotential
\cite{AD}. Besides, in \cite{PK}, a generalized Belinfante procedure
was applied to the canonical conserved quantities in GR \cite{KBL}.
The resulting superpotential is independent of a choice of $g_a$,
and at the same time it coincides only with the superpotential of
the Abbott-Deser type constructed with $l^{\mu\nu}$, see
\cite{Petrov04a}. Thus, in GR the use of the Abbott-Deser type
superpotential with $l^{\mu\nu}$ is more preferable.

However, in {\it arbitrary} gravitation theories, say in $D \neq 4$
dimensions on the AdS backgrounds, we do not know a test model or a
theoretical examination for a choice of perturbations $h_a$. For
example, an arbitrary choice $h_a$ for the solution (\ref{S-AdS})
leads to the same result (\ref{E}) because only the linear order is
crucial. Thus, all the possibilities of the generalized family of
superpotentials constructed with $h^{(a)}_{\alf\beta}$ could be
interesting for future studies.

\bigskip

\noindent{\bf Acknowledgments.} The author is  very grateful to
Nathalie Deruelle for fruitful recommendations that helped to
improve the results and to write the paper in the present form, and
to Joseph Katz for important remarks and discussions. He expresses
his gratitude to Stanley Deser for communication and useful
comments. Author also thanks Deepak Baskaran for a help in checking
English.


\begin{thebibliography}{99}

\bibitem{DT1} Deser S and Tekin B 2002 Gravitational energy in
quadratic curvature gravities {\it Phys. Rev. Lett.} {\bf 89} 101101
({\em Preprint} hep-th/0205318)

\bibitem{DT2} Deser S and Tekin B 2003 Energy in generic higher
curvature gravity theories {\it Phys. Rev.} {\bf D67} 084009 ({\em
Preprint} hep-th/0212292)

\bibitem{DT3} Deser S and Tekin B 2003 Energy in topologically
massive gravity {\it Class. Quantum Grav.} {\bf 20} L259 ({\em
Preprint} gr-qc/0307073)

\bibitem{AD} Abbott L F and Deser S  1982 Stability
of gravity with a cosmological constant {\it Nucl. Phys.} {\bf B195}
76

\bibitem{Paddila} Paddila A 2003 Surface terms and
Gauss-Bonnet Hamiltonian {\it Class. Quantum Grav.} {\bf 20} 3129
({\em Preprint} gr-qc/0303082)

\bibitem{Rubakov} Rubakov V A 2001 Large infinite extra dimensions
{\it Uspekhi Fiz. Nauk} {\bf 171}, 913

\bibitem{Nojiri} Nojiri S, Odintsov S D and  Ogushi S 2002
Friedmann-Robertson-Walker brane cosmological equations from the
five-dimensional bulk (A)dS black hole {\it Int. J. Mod. Phys.} {\bf
A17} 4809 ({\em Preprint} hep-th/0205187)

\bibitem{Der}  Deruelle N and Madore J 2003 On the quasi-linearity of
the Einstein `Gauss-Bonnet' gravity field equations ({\em Preprint}
gr-qc/0305004)

\bibitem{branaP1} Deruelle N,  Dolezel T and  Katz J 2001
Perturbations of brane worlds {\it Phys. Rev.} {\bf D63} 083513
({\em Preprint} hep-th/0010215)

\bibitem{DerKatzOgushi}  Deruelle N, Katz J and  Ogushi S
2004 Conserved charges in Einstein Gauss-Bonnet theory {\it Class.
Quantum Grav.} {\bf 9}, 2521 ({\em Preprint} gr-qc/0310098)

\bibitem{DerKatz} Deruelle N and   Katz J 2005 On the mass of a
Kerr-anti-de Sitter spacetime in $D$ dimensions {\it Class. Quantum
Grav.} {\bf 22} 421 ({\em Preprint} gr-qc/0411035)

\bibitem{PP88} Popova A D and  Petrov A N 1988 The  dynamic  theories  on  a fixed
background in gravitation {\it Int. J. Mod. Phys. }  {\bf A3}  2651

\bibitem{LL} Landau L D and Lifshitz E M
  {\it The Classical Theory of Fields}
  (Pergamon Press, Oxford, 1975).

\bibitem{Mitz} Mitzkevich N V  {\it The Physical Fields in
General Theory of Relativity} (Nauka, Moscow, 1969), in Russian.

\bibitem{Petrov04} Petrov A N 2004 Conserved currents in
$D$-dimensional gravity and brane cosmology{\it Vestnik Mosk.
Univer. Ser. 3. Fiz. Astron.} {\bf No. 2} 10 [2004 {\it Mosc. Univ.
Phys. Bull.} {\bf 59} no. 2   11] ({\em Preprint} gr-qc/0401085)

\bibitem{PK} Petrov A N and Katz J Conserved currents, superpotentials
and cosmological perturbations 2002 {\it Proc. R. Soc. London} {\bf
A458} 319 ({\em Preprint} gr-qc/9911025)

\bibitem{Desercom} Deser S, Kanik I and Tekin B
Conserved charges of higher $D$ Kerr-AdS spacetimes ({\em Preprint}
gr-qc/0506057), and  private communication with Prof. S Deser.

\bibitem{BD+}
Boulware D C and Deser S 1985 String-generated gravity models {\it
Phys. Rev. Lett.} {\bf 55} 2656

\bibitem{BD}
Boulware D C and Deser S 1975 Classical general relativity derived
from quantum gravity {\it Ann. Phys.} {\bf 89} 193


\bibitem{Petrov04a} Petrov A N 2004 Perturbations in the Einstein
theory of gravity: Conserved currents {\it Vestnik Mosk. Univer.
Ser. 3. Fiz. Astron.} {\bf No. 1} 18 [2004 {\it Mosc. Univ. Phys.
Bull.} {\bf 59} no. 1   24] ({\em Preprint} gr-qc/0402090)

\bibitem{BMS} Bondi H, van der Burg M G and Metzner A W K 1962
Gravitational waves in general relativity VII. Waves from
axi-symmetric isolated systems {\it Proc. R. Soc. London} {\bf A269}
21

\bibitem{KBL} Katz J, Bi\v c\'ak J and Lynden-Bell D 1997
Relativistic conservation laws and integral constraints for large
cosmological perturbations {\it Phys. Rev.} {\bf D55}, 5957 ({\em
Preprint} gr-qc/0504041)

\end{thebibliography}
\end{document}